\documentclass[a4paper,11pt]{article}
\usepackage{pos}
\usepackage{multirow}
\usepackage{caption}
\usepackage{subcaption}
\title{Analytic Evaluation of Multiple Mellin-Barnes Integrals}

\author*[a,b]{Sumit Banik}
\author[c,d]{Samuel Friot}

\affiliation[a]{Physik-Institut, Universitat Zurich,\\ Winterthurerstrasse 190, CH 8057 Zurich, Switzerland}

\affiliation[b]{Paul Scherrer Institut, CH 5232 Villigen PSI, Switzerland}

\affiliation[c]{Universit\'e Paris-Saclay, CNRS/IN2P3, \\ IJCLab, 91405 Orsay, France}

\affiliation[d]{Univ Lyon, Univ Claude Bernard Lyon 1, CNRS/IN2P3, \\
 IP2I Lyon, UMR 5822, F-69622, Villeurbanne, France}

\emailAdd{sumit.banik@physik.uzh.ch}
\emailAdd{samuel.friot@universite-paris-saclay.fr}

\abstract{We summarize two geometrical approaches to analytically evaluate higher-fold Mellin-Barnes (MB) integrals in terms of hypergeometric functions. The first method is based on intersections of conic hulls, while the second one, which is more recent, relies on triangulations of a set of points. We demonstrate that, once automatized, the triangulation approach is computationally more efficient than the conic hull approach. As an application of this triangulation approach, we describe how one can derive simpler hypergeometric solutions of the conformal off-shell massless two-loop double box and one-loop hexagon Feynman integrals than those previously obtained from the conic hull approach. Lastly, by applying the above techniques on the MB representation of multiple polylogarithms, we show how to obtain new convergent series representations for these functions. These new analytic expressions were numerically cross-checked with \texttt{GINAC}.
}

\FullConference{Loops and Legs in Quantum Field Theory (LL2024)\\
 14-19, April, 2024\\
Wittenberg, Germany\\}

\begin{document}

\begin{flushright}
PSI-PR-24-16, ZU-TH 36/24
\end{flushright}
\begin{flushright}
\end{flushright}
\vspace{-3cm}
\maketitle

\section{Introduction}
In quantum field theory, theoretical predictions for scattering experiments require computing $S$-matrix elements, which, in the perturbative framework, boils down to computing multi-loop, multi-scale Feynman integrals. As we aim to make theoretical predictions more precise, the number of Feynman integrals that need to be calculated increases significantly. The complexity of individual Feynman integrals also increases drastically because of the higher number of loops, scales, and propagators involved. To solve this problem, an improvement of the existing techniques to evaluate Feynman integrals is essential to achieve breakthroughs in higher-order computations.

Furthermore, beyond the strict boundaries of precision calculation, 
Feynman integrals also play an important role in pure mathematics. For instance, these integrals can be studied using different approaches which involve various branches of mathematics such as the theory of hypergeometric and other special functions~\cite{Kalmykov:2020cqz,Bourjaily:2022bwx,Passarino:2016zcd, Passarino:2024ugq}, GKZ partial differential equations~\cite{delaCruz:2019skx,Klausen:2019hrg,Feng:2019bdx,Ananthanarayan:2022ntm}, Mellin-Barnes integrals \cite{Smirnov:2012gma, Dubovyk:2022obc}, algebraic geometry \cite{Fevola:2023kaw}, etc.\footnote{We only give a few references, many other relevant articles can be found in the literature.} Consequently, advances in computing state-of-the-art Feynman integrals can offer new insights, concepts, and results in some of these branches of mathematics through the perspective of Feynman integrals. 

In this proceeding contribution, we summarize two geometrical approaches to compute some classes of scalar Feynman integrals using their Mellin-Barnes (MB) representation. In the MB approach (see~\cite{Smirnov:2012gma, Dubovyk:2022obc} for an overview), one first converts the Feynman integral into an MB integral (using automated packages~\cite{Gluza:2007rt, Ambre,Belitsky:2022gba}), which yields a more tractable mathematical expression of the Feynman integral. An important breakthrough of this technique, in the context of Feynman integrals, finds its origins in the derivation of the first analytic results for the two-loop box integrals in the planar~\cite{Smirnov:1999gc} and nonplanar~\cite{Tausk:1999vh} cases, followed, among others, by planar master integrals entering Bhabha scattering~\cite{Czakon:2005gi}. More recently, among other notable applications, it has also been used to compute the full two-loop electroweak corrections to the $Z$-boson production and decay~\cite{Dubovyk:2018rlg}, NNLO corrections to Bhabha scattering~\cite{Banerjee:2021mty}, two-loop electroweak corrections to Higgs boson pair production~\cite{Davies:2022ram}, Higgs boson production cross-section at hadronic colliders via gluon fusion at three loops~\cite{Anastasiou:2013srw}, etc. Furthermore, MB integrals are also used in low-energy physics \cite{Charles:2017snx,Ananthanarayan:2017qmx}.

However, certain obstacles currently limit the effectiveness of the MB approach in directly computing Feynman integrals of the most general form. Nevertheless, MB integrals are still very useful for computing the boundary conditions in the differential equation approach and for the asymptotic analysis of Feynman integrals using the method of regions, which can yield parametric integrals independent of kinematical variables where MB integrals are the only tool that can be used \cite{Belitsky:2022gba}.

The outline of this proceeding is as follows. In Section \ref{FI_MB}, we discuss how to derive MB representations for scalar Feynman integrals. In Section \ref{CH_Tri}, we summarize recent geometrical approaches in computing MB integrals in terms of hypergeometric functions and present new solutions for the conformal massless double box and hexagon Feynman integrals. Section \ref{MB_MPLs} is dedicated to evaluating the MB integral representations of multiple polylogarithms. Finally, in Section \ref{Conclusion}, we provide concluding remarks and future directions for further exploration.

\section{Feynman Diagrams and MB integrals}
\label{FI_MB}
The link between Feynman integrals and hypergeometric functions was first established in Ref.~\cite{Regge} almost sixty years ago. Therefore, this indirectly relates Feynman integrals with MB integrals, since it is known for over a century that many multivariable hypergeometric functions have an MB representation. At present, the link between Feynman integrals and MB integrals is more direct, and it relies on the simple formula
\begin{equation}
     \frac{1}{(A+B)^{\alpha}}=\frac{1}{\Gamma(\alpha)}\int_{-i \infty}^{+i \infty} \frac{d z}{2 \pi i} \Gamma(-z) \Gamma(\alpha+z) A^{-\alpha-z} B^{z}
     \label{MB_identiy}
\end{equation}

which can be recursively applied on each propagator in the momentum representation of a given Feynman diagram to yield a MB representation. Alternatively, if the Feynman integral is written in the Schwinger parameter representation, then Eq.~\eqref{MB_identiy} can be applied to the corresponding $U$ and $F$ polynomials to obtain the MB representation. One can also start from the Feynman parameter representation and these approaches were widely used during the 1990s and early 2000s (see~\cite{Smirnov:2012gma} and the references therein) until they were automated in the \textit{Mathematica} package \texttt{AMBRE}~\cite{Gluza:2007rt, Ambre}. For a recent alternative package, see also \texttt{MBcreate}~\cite{Belitsky:2022gba}.
 
As an illustration, we show below (up to some overall factor) the 3-fold MB representation of the off-shell one-loop three-point scalar Feynman integral with three equal masses~\cite{Davydychev:1990jt}:
\begin{align}
	&\int\limits_{-i \infty}^{+i \infty}\frac{dz_1}{2\pi i}\int\limits_{-i \infty}^{+i \infty}\frac{dz_2}{2\pi i}\int\limits_{-i \infty}^{+i \infty}\frac{dz_3}{2\pi i}
	(-X_1)^{z_1}(-Y_1)^{z_2}(-Z_1)^{z_3}\Gamma(-z_1)\Gamma(-z_2)\Gamma(-z_3)\nonumber\\
	&\times \Gamma(\nu_1+z_1+z_2)\Gamma(\nu_2+z_2+z_3)\Gamma(\nu_3+z_1+z_3)
	\frac{\Gamma(\nu_1+\nu_2+\nu_3-n/2+z_1+z_2+z_3)}{\Gamma(\nu_1+\nu_2+\nu_3+2z_1+2z_2+2z_3)}
\end{align}
where the $\nu_i$ $(i=1,2,3)$ are the powers of propagators, $n$ is the space-time dimension, and we refer the reader to~\cite{Davydychev:1990jt} for the precise forms of $X_1, Y_1, Z_1$ which are combinations of the squared external momentum and the mass. For more details on MB integrals and their applications in quantum field theory, we refer the reader to the recent book~\cite{Dubovyk:2022obc}.  We next summarize the recent developments in evaluating multi-fold MB integrals analytically.

\section{Evaluating multiple MB integrals analytically}
\label{CH_Tri}
A general $N$-fold MB integral can be expressed as \begin{align}
  I_N=\int\limits_{-i \infty}^{+i \infty} \frac{ d z_1}{2 \pi i} \cdots \int\limits_{-i \infty}^{+i \infty}\frac{ d z_N}{2 \pi i}\,\, \,  x^{z_1}_{1} \cdots x^{z_N}_{N} \, \, \, \frac{\prod\limits_{i=1}^{k} \Gamma^{a_i}({\bf e}_i\cdot{\bf \mathbf{z}}+g_i)}{\prod\limits_{j=1}^{l}\Gamma^{b_j}({\bf f}_j\cdot{\bf z}+h_j)}\label{NfoldMB}
  \end{align}
  where $a_i , b_j, k$ and $l$ are positive integers, ${\bf e}_i$, ${\bf f}_j$ are $N$-dimensional coefficient real vectors, ${\bf z}=(z_1, ... , z_N)$ and $g_i, h_j$ can be complex numbers. When not specified, the contours of integration are implicitly chosen such that they do not separate the set of poles of each of the numerator gamma functions into different subsets. This implies that for some cases the contours may not be straight lines parallel to the imaginary axis of the integration variables.

The evaluation of one-fold MB integrals is straightforward and is described in detail in several books such as~\cite{Marichev, Paris}.
However, the multifold case is much more complicated.
Therefore, there have been many efforts to develop automated tools for computing MB representations in the context of the Feynman diagram calculations \cite{MBtools}. For instance, the \textit{Mathematica} package \texttt{MBsums.m}~\cite{Ochman:2015fho} was created a few years ago for their analytic evaluation, but it is limited to MB integrals with straight contours and the analytic results derived from it are often bulky due to the appearance of spurious terms.

In parallel to the above developments, some mathematicians worked on a non-iterative and rigorous approach to evaluate MB integrals based on the theory of multidimensional residues~\cite{Tsikh1}, as detailed in a series of papers~\cite{Tsikh2, Passare:1996db, Tsikh3}. However, to the best of our knowledge, they did not explicitly provide the formulas necessary for evaluating $N$-fold MB integrals when $N>2$, even in the non-logarithmic (or non-resonant) case\footnote{For $N=2$, the logarithmic case was considered for MB integrals with straight contours in~\cite{Friot:2011ic}.}. Inspired by their work and with the aim of filling this gap, we developed a new and efficient method based on conic hulls and their intersections~\cite{Ananthanarayan:2020fhl}. Although not established with the same level of rigor as in previous works~\cite{Tsikh2, Passare:1996db, Tsikh3}, this method has been tested on many examples, either analytically when results were available in the literature or numerically using sector decomposition~\cite{Smirnov:2021rhf,Borowka:2017idc}. In the following, we will describe this method in more detail, but first, we recall a few general features about MB integrals.

MB integrals can be divided into two main categories (see \cite{Tsikh3} for more details): 
\begin{itemize}
     \item The \textit{degenerate} type, where ${\bf \Delta}\doteq\sum a_i  {\bf e_i} - \sum b_j {\bf f_j}={\bf 0}$ (see Eq.~\eqref{NfoldMB}). In this case, several hypergeometric series solutions of the corresponding MB integral coexist, which are, as a rule, analytic continuations of each other. It is observed that the MB representations of all scalar Feynman integrals fall into this class.
 
     \item The \textit{non-degenerate} type, which satisfies ${\bf\Delta} \neq {\bf 0}$. In this case, one or more convergent series representations converge for all values of $x_1 \cdots x_N$. In addition, there will be asymptotic series solutions.
 \end{itemize}
 Based on the singularity structure of the MB integrand, they can be further classified into two sub-types:
  \begin{itemize}
      \item  The \textit{non-resonant} case: Here, all the poles of the MB integrand are of order one. In the context of Feynman integrals, the MB representation usually falls into this category if the powers of the propagators are generic.
      
      \item The \textit{resonant} or the \textit{logarithmic} case. Here, some or all poles of the MB integral are of order greater than one. Therefore, residue computation leads to logarithmic solutions. In the context of Feynman integrals, this typically happens when the powers of the propagators are set to integers.
  \end{itemize} 

In the following, we describe the conic hull and triangulation methods to compute $N$-fold MB integrals. Both of these methods can be used to compute MB integrals that fall into any of the above cases.

\subsection{Brief overview of the conic hull method (non-resonant case)}

We briefly outline the main steps of the conic hull in the non-resonant case. For more details on the technical aspects and the resonant case, we refer the reader to~\cite{Ananthanarayan:2020fhl}.

\begin{itemize}
\item \textit{Step 1:} For a given $N$-fold MB integral, find all possible $N$-combinations of the numerator gamma functions and retain only the non-singular ones.

\item \textit{Step 2:} Associate a series, which we call a \textit{building block}, with each retained combination.

\item \textit{Step 3:} Construct a conic hull for each combination/building block.

\item \textit{Step 4:} Find the \textit{largest} intersecting subsets of conic hulls. The sum of the building blocks associated with the conic hulls in each of these subsets yields a series representation of the MB integral.

\item \textit{Step 5:} The intersection region of the largest subset of conic hulls is usually a conic hull, which we call the \textit{master conic hull}. Using it, we can derive a master series which considerably simplifies the convergence analysis of the series representation.
\end{itemize}

To make the method ready-to-use, we have automatized it in a  \textit{Mathematica} package called \texttt{MBConicHulls.wl}~\cite{Ananthanarayan:2020fhl, MBConicHullsGit}. The package also works for the resonant case, which is more complicated because it involves multivariate residues. Therefore, to handle the resonant case, we have internally used the \textit{Mathematica} package \texttt{MultivariateResidues.m}~\cite{Larsen:2017aqb}. 
   
The first version of \texttt{MBConicHulls.wl} only worked for MB integrals with non-straight contours separating the poles of the integrand as described above. It has been successfully applied in the computation of various unsolved non-resonant and resonant MB representations of Feynman integrals~\cite{Ananthanarayan:2020ncn, Ananthanarayan:2020xpd,Datta:2023otd}. Subsequently, we focused on the case of arbitrary straight contours of integration (parallel to the imaginary axis in the complex planes of the integration variables) since these integrals frequently appear if one performs an $\epsilon$-expansion of the MB integral using \texttt{MB.m}~\cite{Czakon:2005rk} or \texttt{MBresolve.m}~\cite{Smirnov:2009up}. We solved the case of straight contours in~\cite{Banik:2022bmk} and implemented it in a new version of \texttt{MBConicHulls.wl}~\cite{MBConicHullsGit}. The solution to this problem relies on the appropriate use of the generalized reflection formula

  \begin{align*}
  \Gamma(z-n)=\frac{\Gamma(z)\Gamma(1-z)(-1)^{n}}{\Gamma(n+1-z)} \,\, \text{for } n \in \mathbb{Z}
  \end{align*}
 in such a way that the real part of each numerator gamma function of the MB integrand becomes positive along the integration contours. The conic hull method can then be applied in the same way as it is for non-straight contours.

\subsection{The triangulation method}
     
Although the conic hull method is very efficient, it is often too slow when dealing with (very) complicated higher-fold MB integrals. In trying to improve the method, we found that regular triangulations of a particular point configuration are dual to
the relevant intersections of conic hulls and can thus be used to derive different series representations of a given MB integral~\cite{Banik:2023rrz}. Although the two approaches would yield the same set of series solutions, the great advantage of triangulations is that once automatized, they can provide the results for complicated integrals much faster than conic hulls.
We have thus recently upgraded the \texttt{MBConicHulls.wl} package such that, in addition to the original analysis based on conic hulls, it offers the possibility of computing MB integrals using triangulations. This is possible by internally using the software \texttt{TOPCOM}~\cite{Rambau:TOPCOM:2002} to perform triangulations. Therefore, one is now able to solve much more complicated MB integrals than before.

In general, to apply the triangulation method, we need to perform a change of the integration variables to rewrite Eq. \eqref{NfoldMB} in the \textit{canonical form}
\begin{equation} \label{N_MB_2}
    I_N= \int\limits_{-i \infty}^{+i \infty} \frac{ d z_1}{2 \pi i} \cdots \int\limits_{-i \infty}^{+i \infty}\frac{ d z_N}{2 \pi i}\,\,  \frac{ \Gamma(-z_1)\cdots\Gamma(-z_N) \prod\limits_{i=N+1}^{k'} \Gamma^{a'_i}(s'_i ({\bf z})) }{\prod\limits_{j=1}^{l} \Gamma^{b'_j}(t'_j ({\bf z}))} x'^{z_1}_{1} \cdots x'^{z_N}_{N}
\end{equation}
where we have pulled out the factors $ \Gamma(-z_1)\cdots\Gamma(-z_N) $ in the numerator for convenience. The coefficient vectors $s'_i$ and $t'_j$ are expressed as follows:
\begin{align}\label{argument_canonical}
    s'_i({\bf z}) =\sum\limits_{k=1}^{N}e'_{i k}z_k+f'_i \, , \hspace{2cm}
    t'_j({\bf z}) =\sum\limits_{k=1}^{N}g'_{j k}z_k+h'_j
\end{align}
The next step is to construct the point configuration on which we perform triangulations. The configuration has a set of $N$ points\footnote{Each of these $N$ points has $\sum_{i=N+1}^{k'}a_i'$ coordinates as we include the possibility of non-unit positive powers of the numerator gamma functions.}
  \begin{equation*}
P_1 = 
e'_{l 1} \,\, , \hspace{1cm}
P_2 = 
e'_{l 2} \,\, , \hspace{0.5cm} \cdots \hspace{0.5cm}
P_{N} = 
e'_{l N}
\end{equation*}
and $\sum_{i=N+1}^{k'}a_i'$ additional points corresponding to the unit vectors of dimension $\sum_{i=N+1}^{k'}a_i'$. The complete set of these $N+\sum_{i=N+1}^{k'}a_i'$ points is then used for triangulations, performed by calling \texttt{TOPCOM} internally from \texttt{MBConicHulls.wl}.

\begin{figure}[htb]
     \centering
     \begin{subfigure}[b]{0.3\textwidth}
         \centering
         \includegraphics[scale=1.4]{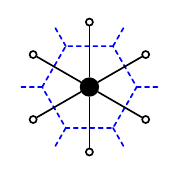}
         \caption{Conformal hexagon}
     \label{Conformal_HX_Diagram}
     \end{subfigure}
     \begin{subfigure}[b]{0.3\textwidth}
         \centering
         \includegraphics[scale=1.4]{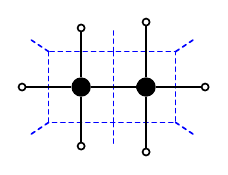}
         \caption{Conformal double-box}
        \label{Conformal_DB_Diagram}
     \end{subfigure}
     \hspace{1cm}
     \begin{subfigure}[b]{0.3\textwidth}
         \includegraphics[scale=1.2]{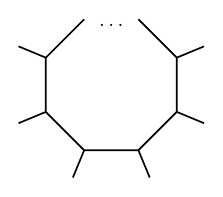}
         \caption{One-Loop $N$-point Massless}
        \label{NPoint_Diagram}
     \end{subfigure}
        \caption{Some of the one-loop and two-loop Feynman diagrams evaluated using the conic hull and triangulation methods in Table \ref{Speed}.}
        \label{Feynman_Diagrams}
\end{figure}

We applied the triangulation method to various Feynman integrals with higher-order MB representations. For example, we recomputed the massless conformal double-box and hexagon Feynman diagrams in Fig.~\ref{Feynman_Diagrams}, originally solved using the conic-hull approach~\cite{Ananthanarayan:2020ncn}. Thanks to the efficiency of the triangulation approach, we could obtain all possible series representations that the method could offer. Due to this, we could derive new series solutions for both conformal integrals as a sum of 25 hypergeometric series, surpassing the results previously obtained in \cite{Ananthanarayan:2020ncn}.
To further test the package, we computed the off-shell one-loop massless $N$-point Feynman integral in Fig.~\ref{Feynman_Diagrams} with generic powers of the propagators in order to gauge the computational power of the triangulation approach. We show in Table~\ref{Speed} the speed comparison of the triangulation and conic hull approaches.

\begin{table}[h]
  \centering
  \renewcommand{\arraystretch}{1.3}
  \begin{tabular}{|p{2cm}|p{0.7cm}|p{1.4cm}|c|c|c|c|}
    \hline
    {Feynman integral} & {MB folds} & \multirow{2}{1.5cm}{Total solution number}  & \multicolumn{2}{c|}{Conic hull method} & \multicolumn{2}{c|}{Triangulation method} \\[0.3cm]
    \cline{4-7}
  & &  & One solution & All solutions & One solution & All solutions \\
    \hline
    Conformal triangle & 3 & 14 & 0.186 sec. & 1.44 sec. & 0.205 sec.  & 0.483 sec.  \\[0.3cm] \hline
    Massless pentagon & 5 & 70  & 1.276 sec.  & 1.25 h. & 0.318 sec. & 2.78 sec.  \\[0.3cm] \hline
    Conformal hexagon & 9 & 194160 & 1 min. & -  & 0.489 sec. & 40 min. \\[0.3cm] \hline
    Conformal double-box & 9 & 243186 & 1.9 min. & - & 0.635 sec. & 1.8 h.\\[0.3cm] \hline
    Hard diagram & 8 & 1471926 & 6 min. & - & 1.4 sec. & - \\[0.3cm] \hline
  \end{tabular}
  \caption{Speed comparison of the conic hull and triangulation methods. For more details, see Ref.~\cite{Banik:2023rrz}.}
  \label{Speed}
\end{table}

\section{Mellin-Barnes representation of Multiple Polylogarithms}
\label{MB_MPLs}
Multiple polylogarithms (MPLs) are presently an important class of functions in mathematics that appear in many diverse areas of modern high-energy physics. In particular, these functions occur in almost every modern multi-loop calculation. Therefore, further exploration of the mathematical properties of these functions is worthwhile. They can be defined in terms of nested sums as
\begin{align}
\text{Li}_{m_1,\cdots, m_n}(x_1, \cdots, x_n)&=\sum_{0<k_1<k_2<\cdots<k_n}^{\infty}\frac{x_1^{k_1} x_2^{k_2} \cdots x^{k_n}_n}{k_1^{m_1}k_2^{m_2}  \cdots k^{m_n}_n}
\end{align}
Here, we focus on studying the relatively unexplored MB representation~\cite{Anastasiou:2013mca} of these functions, which can also be expressed as
\begin{align}
&\text{Li}_{m_1,...,m_N}(x_1,\cdots ,x_N) =x_1 x^2_2 \cdots x^N_N \int\limits_{-i \infty}^{+i \infty} \frac{ d z_1}{2 \pi i} \cdots\int\limits_{-i \infty}^{+i \infty} \frac{ d z_N}{2 \pi i} (-x_1 x_2 \cdots x_N)^{z_1} (- x_2 \cdots x_N)^{z_2}\cdots (-x_N )^{z_N}\nonumber \\ & \times\Gamma(-z_1)\cdots\Gamma(-z_N)\Gamma(1+z_1)\cdots\Gamma(1+z_N) 
\frac{ \Gamma^{m_1}(1+z_1)\Gamma^{m_2}(2+z_{12})}{\Gamma^{m_1}(2+z_1)\Gamma^{m_2}(3+z_{12})} \cdots  \frac{\Gamma^{m_N}(N+z_{1\cdots N})}{\Gamma^{m_N}(N+1+z_{1\cdots N})}
\end{align}
We applied the conic hull and triangulation approaches on various particular cases of this multifold MB integral in order to derive convergent hypergeometric series solutions \cite{wip}. As these MB integrals are of degenerate type, we obtain in each case several series representations which are analytic continuations of one another and, to the best of our knowledge, are mostly unknown in the literature. We validated these new expressions numerically using the \texttt{G\small{I}N\small{A}C}~\cite{Bauer:2000cp} interface of the  \texttt{PolyLogTools}~\cite{Duhr:2019tlz} package (see also  \cite{Vollinga:2004sn}).

\begin{figure}[htb!]
    \centering
    \includegraphics[scale=0.55]{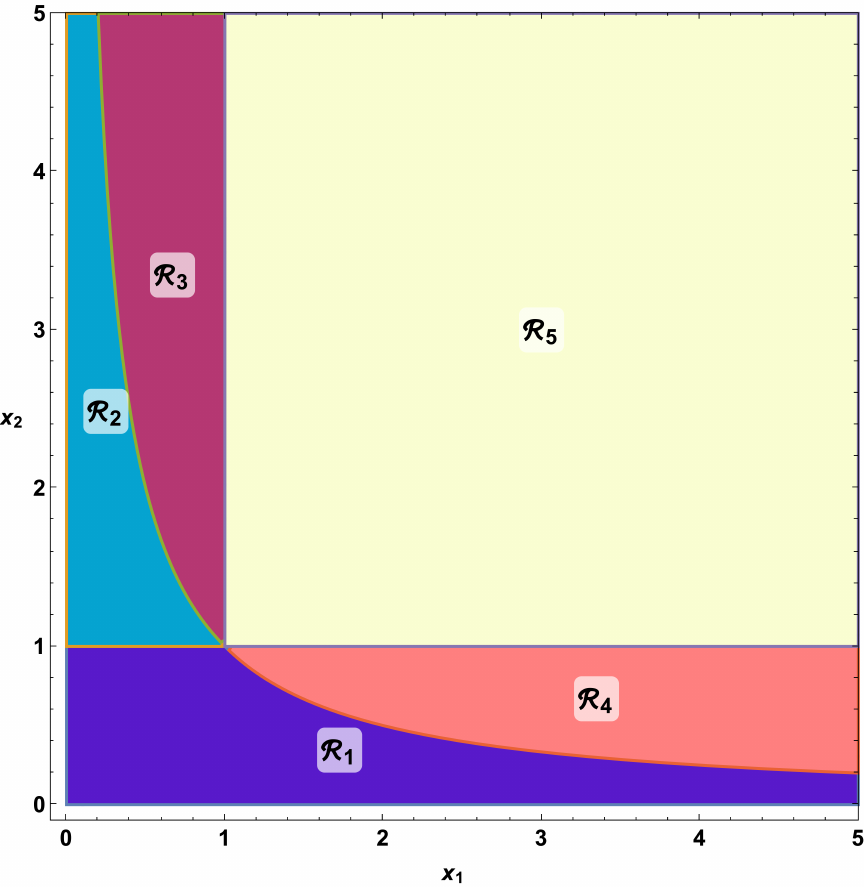}
    \caption{Regions of convergence of the five series representations of the $2$-fold MB representation of $\text{Li}_{m_1,m_2}(x_1,x_2)$.}
    \label{ROC}
\end{figure}

While performing the above exercise, we also observed that MPLs belong to a special class of MB integrals having no \textit{white zones}. This means that, using either the conic hull or triangulation approach, we can derive a set of series solutions converging for nearly all possible values of $x_1, \cdots, x_N $. For example, we obtained five series solutions for the MB representation of $\text{Li}_{m_1,m_2}(x_1,x_2)$ using the conic-hull approach, which is enough to fill the whole $(x_1, x_2)$ region (except on the boundaries of the $\mathcal{R}_i$), as shown in Fig.~\ref{ROC}. Note that this is true for any values of the depths $m_1$ and $m_2$ in $\text{Li}_{m_1,m_2}(x_1,x_2)$.

\section{Conclusions}
\label{Conclusion}
We have shown that $N$-fold MB integrals can be analytically computed in a non-iterative way using conic hulls or triangulations of point configurations. These methods, which can handle both resonant and non-resonant cases, are automated in the \texttt{MBConicHulls.wl} package. This package was used for the first analytic computation of the massless off-shell conformal hexagon and double box for both generic and unit powers of the propagators. In addition, it can be used to systematically study the transformation theory of multivariate hypergeometric functions, an area that has been largely unexplored. Finally, we derived convergent series representations of MPLs valid for all regions in the parameter space (except for the specific values that belong to the boundaries of the convergence regions of the series representations where additional conditions may have to be considered).

However, some work remains to improve these geometrical methods for applications to phenomenologically relevant precision calculations. For instance, our aim is now to calculate MB representations of Feynman integrals with fewer scales than integration variables. Currently, this is possible in some cases but not all. 
It would also be beneficial to better understand the master series conjecture~\cite{Ananthanarayan:2020fhl} and to find the rigorous link between the geometries of triangulations and conic hulls. Another important issue concerns the “white zones” where none of the series representations derived from the direct evaluation of many MB integrals using the triangulation/conic hull approach converges. Finding a systematic way to obtain series valid in these regions, which can be of phenomenological relevance, would be important.
Finally, a deeper systematic study of nondegenerate MB integrals and their links with asymptotics in several variables, resurgence, and related topics could lead to interesting insights and new results.

\section*{Acknowledgments}

We thank B. Ananthanarayan and S. Ghosh for initial collaboration on the conic hull theory. The work of S.B.~is supported by the Swiss National Science Foundation Grant Number PP00P21\_76884 .


\begin{thebibliography}{99}


\bibitem{Kalmykov:2020cqz}
M.~Kalmykov, V.~Bytev, B.~A.~Kniehl, S.~O.~Moch, B.~F.~L.~Ward and S.~A.~Yost,
doi:10.1007/978-3-030-80219-6\_9
[arXiv:2012.14492 [hep-th]].

\bibitem{Bourjaily:2022bwx}
J.~L.~Bourjaily, J.~Broedel, E.~Chaubey, C.~Duhr, H.~Frellesvig, M.~Hidding, R.~Marzucca, A.~J.~McLeod, M.~Spradlin and L.~Tancredi, \textit{et al.}
[arXiv:2203.07088 [hep-ph]].

\bibitem{Passarino:2016zcd}
G.~Passarino,
Eur. Phys. J. C \textbf{77} (2017) no.2, 77
doi:10.1140/epjc/s10052-017-4623-1
[arXiv:1610.06207 [math-ph]].

\bibitem{Passarino:2024ugq}
G.~Passarino,
[arXiv:2405.18755 [hep-ph]].






\bibitem{delaCruz:2019skx}
L.~de la Cruz,
JHEP \textbf{12} (2019), 123
doi:10.1007/JHEP12(2019)123
[arXiv:1907.00507 [math-ph]].

\bibitem{Klausen:2019hrg}
R.~P.~Klausen,
JHEP \textbf{04} (2020), 121
doi:10.1007/JHEP04(2020)121
[arXiv:1910.08651 [hep-th]].

\bibitem{Feng:2019bdx}
T.~F.~Feng, C.~H.~Chang, J.~B.~Chen and H.~B.~Zhang,
Nucl. Phys. B \textbf{953} (2020), 114952
doi:10.1016/j.nuclphysb.2020.114952
[arXiv:1912.01726 [hep-th]].

\bibitem{Ananthanarayan:2022ntm}
B.~Ananthanarayan, S.~Banik, S.~Bera and S.~Datta,
Comput. Phys. Commun. \textbf{287} (2023), 108699
doi:10.1016/j.cpc.2023.108699
[arXiv:2211.01285 [hep-th]].

\bibitem{Smirnov:2012gma}
V.~A.~Smirnov,
\textit{``Analytic tools for Feynman integrals''},
Springer Tracts Mod. Phys. \textbf{250} (2012), 1-296
doi:10.1007/978-3-642-34886-0

\bibitem{Dubovyk:2022obc}
I.~Dubovyk, J.~Gluza and G.~Somogyi,
\textit{``Mellin-Barnes Integrals: A Primer on Particle Physics Applications,''}
Lect. Notes Phys. \textbf{1008} (2022), pp.
doi:10.1007/978-3-031-14272-7
[arXiv:2211.13733 [hep-ph]].





\bibitem{Fevola:2023kaw}
C.~Fevola, S.~Mizera and S.~Telen,
Phys. Rev. Lett. \textbf{132} (2024) no.10, 101601
doi:10.1103/PhysRevLett.132.101601
[arXiv:2311.14669 [hep-th]].


\bibitem{Gluza:2007rt}
J.~Gluza, K.~Kajda and T.~Riemann,
Comput. Phys. Commun. \textbf{177} (2007), 879-893
doi:10.1016/j.cpc.2007.07.001
[arXiv:0704.2423 [hep-ph]].

\bibitem{Ambre}
AMBRE webpage: http://prac.us.edu.pl/gluza/ambre


\bibitem{Belitsky:2022gba}
A.~V.~Belitsky, A.~V.~Smirnov and V.~A.~Smirnov,
Nucl. Phys. B \textbf{986} (2023), 116067
doi:10.1016/j.nuclphysb.2022.116067
[arXiv:2211.00009 [hep-ph]].

\bibitem{Smirnov:1999gc}
V.~A.~Smirnov,
Phys. Lett. B \textbf{460} (1999), 397-404
doi:10.1016/S0370-2693(99)00777-7
[arXiv:hep-ph/9905323 [hep-ph]].


\bibitem{Tausk:1999vh}
J.~B.~Tausk,
Phys. Lett. B \textbf{469} (1999), 225-234
doi:10.1016/S0370-2693(99)01277-0
[arXiv:hep-ph/9909506 [hep-ph]].

\bibitem{Czakon:2005gi}
M.~Czakon, J.~Gluza and T.~Riemann,
Acta Phys. Polon. B \textbf{36} (2005), 3319-3326
[arXiv:hep-ph/0511187 [hep-ph]].

\bibitem{Dubovyk:2018rlg}
I.~Dubovyk, A.~Freitas, J.~Gluza, T.~Riemann and J.~Usovitsch,
Phys. Lett. B \textbf{783} (2018), 86-94
doi:10.1016/j.physletb.2018.06.037
[arXiv:1804.10236 [hep-ph]].

\bibitem{Banerjee:2021mty}
P.~Banerjee, T.~Engel, N.~Schalch, A.~Signer and Y.~Ulrich,
Phys. Lett. B \textbf{820} (2021), 136547
doi:10.1016/j.physletb.2021.136547
[arXiv:2106.07469 [hep-ph]].

\bibitem{Davies:2022ram}
J.~Davies, G.~Mishima, K.~Sch\"onwald, M.~Steinhauser and H.~Zhang,
JHEP \textbf{08} (2022), 259
doi:10.1007/JHEP08(2022)259
[arXiv:2207.02587 [hep-ph]].

\bibitem{Anastasiou:2013srw}
C.~Anastasiou, C.~Duhr, F.~Dulat and B.~Mistlberger,
JHEP \textbf{07} (2013), 003
doi:10.1007/JHEP07(2013)003
[arXiv:1302.4379 [hep-ph]].

\bibitem{Charles:2017snx}
J.~Charles, E.~de Rafael and D.~Greynat,
Phys. Rev. D \textbf{97} (2018) no.7, 076014
doi:10.1103/PhysRevD.97.076014
[arXiv:1712.02202 [hep-ph]].

\bibitem{Ananthanarayan:2017qmx}
B.~Ananthanarayan, J.~Bijnens, S.~Friot and S.~Ghosh,
Phys. Rev. D \textbf{97} (2018) no.9, 091502
doi:10.1103/PhysRevD.97.091502
[arXiv:1711.11328 [hep-ph]].

\bibitem{Regge} 
T. Regge,  \textit{``Algebraic Topology Methods in the Theory of Feynman Relativistic Amplitudes,''} Battelle Rencontres (1967) Lectures in Mathematics and Physics. Ed. by C. M. De Witt and J. A. Wheeler, New York: Benjamin, 1968, pp. 433-458. 


\bibitem{Davydychev:1990jt}
A.~I.~Davydychev,
J. Math. Phys. \textbf{32} (1991), 1052-1060
doi:10.1063/1.529383

\bibitem{Marichev}
  O. I.~Marichev, 
  \textit{``Handbook of integral transforms of higher transcendental functions: Theory and Algorithmic tables''},
 Ellis Horwood Series in Mathematics and Its Applications, 1983.


\bibitem{Paris}
R. B. Paris and D. Kaminski,
\textit{``Asymptotics and Mellin-Barnes integrals,''} Encyclopedia of Mathematics and its Applications, Vol. \textbf{85}, Cambridge University Press, Cambridge, 2001.

\bibitem{MBtools}
https://mbtools.hepforge.org

\bibitem{Ochman:2015fho}
M.~Ochman and T.~Riemann,
Acta Phys. Polon. B \textbf{46} (2015) no.11, 2117
doi:10.5506/APhysPolB.46.2117
[arXiv:1511.01323 [hep-ph]].

\bibitem{Tsikh1}
A. Tsikh,  \textit{``Multidimensional residues and their applications,''}  [in Russian] Nauka, Novosibirsk.
(1988).

\bibitem{Tsikh2}
M. Passare, A. Tsikh, and O. Zhdanov, 
Aspects Math., 233-241 (1994).

\bibitem{Passare:1996db}
M.~Passare, A.~K.~Tsikh and A.~A.~Cheshel,
Teor. Mat. Fiz. \textbf{109N3} (1996), 381-394
doi:10.1007/BF02073871
[arXiv:hep-th/9609215 [hep-th]].


\bibitem{Tsikh3}
A. Tsikh, O. Zhdanov, 
Siberian Math. J., 39(1998), no. 2, 245-260.

\bibitem{Friot:2011ic}
S.~Friot and D.~Greynat,
J. Math. Phys. \textbf{53} (2012), 023508
doi:10.1063/1.3679686
[arXiv:1107.0328 [math-ph]].

\bibitem{Ananthanarayan:2020fhl}
B.~Ananthanarayan, S.~Banik, S.~Friot and S.~Ghosh,
Phys. Rev. Lett. \textbf{127} (2021) no.15, 151601
doi:10.1103/PhysRevLett.127.151601
[arXiv:2012.15108 [hep-th]].



\bibitem{Smirnov:2021rhf}
A.~V.~Smirnov, N.~D.~Shapurov and L.~I.~Vysotsky,
Comput. Phys. Commun. \textbf{277} (2022), 108386
doi:10.1016/j.cpc.2022.108386
[arXiv:2110.11660 [hep-ph]].

\bibitem{Borowka:2017idc}
S.~Borowka, G.~Heinrich, S.~Jahn, S.~P.~Jones, M.~Kerner, J.~Schlenk and T.~Zirke,
Comput. Phys. Commun. \textbf{222} (2018), 313-326
doi:10.1016/j.cpc.2017.09.015
[arXiv:1703.09692 [hep-ph]].


\bibitem{MBConicHullsGit}
\texttt{MBConicHulls.wl} webpage: https://github.com/SumitBanikGit/MBConicHulls

\bibitem{Larsen:2017aqb}
K.~J.~Larsen and R.~Rietkerk,
Comput. Phys. Commun. \textbf{222} (2018), 250-262
doi:10.1016/j.cpc.2017.08.025
[arXiv:1701.01040 [hep-th]].

\bibitem{Ananthanarayan:2020ncn}
B.~Ananthanarayan, S.~Banik, S.~Friot and S.~Ghosh,
Phys. Rev. D \textbf{102} (2020) no.9, 091901
doi:10.1103/PhysRevD.102.091901
[arXiv:2007.08360 [hep-th]].

\bibitem{Ananthanarayan:2020xpd}
B.~Ananthanarayan, S.~Banik, S.~Friot and S.~Ghosh,
Phys. Rev. D \textbf{103}, no.9, 096008 (2021)
doi:10.1103/PhysRevD.103.096008
[arXiv:2012.15646 [hep-th]].




\bibitem{Datta:2023otd}
S.~Datta, N.~Rana, V.~Ravindran and R.~Sarkar,
JHEP \textbf{12} (2023), 001
doi:10.1007/JHEP12(2023)001
[arXiv:2308.12169 [hep-ph]].

\bibitem{Czakon:2005rk}
M.~Czakon,
Comput. Phys. Commun. \textbf{175} (2006), 559-571
doi:10.1016/j.cpc.2006.07.002
[arXiv:hep-ph/0511200 [hep-ph]].

\bibitem{Smirnov:2009up}
A.~V.~Smirnov and V.~A.~Smirnov,
Eur. Phys. J. C \textbf{62} (2009), 445-449
doi:10.1140/epjc/s10052-009-1039-6
[arXiv:0901.0386 [hep-ph]].


\bibitem{Banik:2022bmk}
S.~Banik and S.~Friot,
Phys. Rev. D \textbf{107} (2023) no.1, 016007
doi:10.1103/PhysRevD.107.016007
[arXiv:2212.11839 [hep-ph]].

\bibitem{Banik:2023rrz}
S.~Banik and S.~Friot, accepted in Phys. Rev. D
[arXiv:2309.00409 [hep-th]].

\bibitem{Rambau:TOPCOM:2002}
J. Rambau, \textit{``TOPCOM: Triangulations of Point Configurations and Oriented Matroids,''}
 Mathematical Software - ICMS 2002 (Cohen, Arjeh M. and Gao, Xiao-Shan and Takayama, Nobuki, eds.), World Scientific (2002), pp. 330-340.

\bibitem{Anastasiou:2013mca}
C.~Anastasiou, C.~Duhr, F.~Dulat, F.~Herzog and B.~Mistlberger,
JHEP \textbf{12} (2013), 088
doi:10.1007/JHEP12(2013)088
[arXiv:1311.1425 [hep-ph]].

\bibitem{wip}
S.~Banik and S.~Friot, work in progress.




\bibitem{Bauer:2000cp}
C.~W.~Bauer, A.~Frink and R.~Kreckel,
J. Symb. Comput. \textbf{33} (2002), 1-12
doi:10.1006/jsco.2001.0494
[arXiv:cs/0004015 [cs.SC]].



\bibitem{Duhr:2019tlz}
C.~Duhr and F.~Dulat,
JHEP \textbf{08} (2019), 135
doi:10.1007/JHEP08(2019)135
[arXiv:1904.07279 [hep-th]].

\bibitem{Vollinga:2004sn}
J.~Vollinga and S.~Weinzierl,
Comput. Phys. Commun. \textbf{167} (2005), 177
doi:10.1016/j.cpc.2004.12.009
[arXiv:hep-ph/0410259 [hep-ph]].



%
























\end{thebibliography}
\end{document}